\documentclass[prd,twocolumn,nofootinbib]{revtex4-1}
\usepackage{amsfonts}
\usepackage{amsmath}
\usepackage{amssymb}
\usepackage{bm}
\usepackage{dcolumn}
\usepackage[dvips]{graphicx}
\usepackage{graphics}
\usepackage{latexsym}
\usepackage{rotating}
\usepackage[colorlinks=true]{hyperref}
\usepackage{xspace} 
\usepackage[usenames]{color}
\usepackage{mathrsfs}
\usepackage{multirow}
\usepackage{pifont}
\usepackage{enumitem}
\usepackage{color}
\usepackage{appendix}
\usepackage[utf8]{inputenc}
\usepackage{comment}
\definecolor {darkgreen}{rgb}{0.2,0.7,0.2}
\definecolor{purple}{rgb}{0.5,0,0.5}

\newcommand\be{\begin{equation}}
\newcommand\ba{\begin{eqnarray}}
\newcommand\ee{\end{equation}}
\newcommand\ea{\end{eqnarray}}
\newcommand\bw{\begin{widetext}}
\newcommand\ew{\end{widetext}}

\newcommand{\EDGB}{{\mbox{\tiny EdGB}}}
\newcommand{\BD}{{\mbox{\tiny BD}}}
\newcommand{\NC}{{\mbox{\tiny NC}}}

\newcommand{\KG}{{\mbox{\tiny kh}}}
\newcommand{\EA}{{\mbox{\tiny EA}}}
\newcommand{\ST}{{\mbox{\tiny ST}}}

\newcommand{\DCS}{{\mbox{\tiny dCS}}}
\newcommand{\GR}{{\mbox{\tiny GR}}}
\newcommand{\Gdot}{{\mbox{\tiny $\dot G$}}}
\newcommand{\GW}{{\mbox{\tiny GW}}}

\begin{document}
\title{Parameterized Post-Einsteinian Gravitational Waveforms in \\ Various Modified Theories of Gravity}

\author{Sharaban Tahura}
\affiliation{Department of Physics, University of Virginia, Charlottesville, Virginia 22904, USA.}

\author{Kent Yagi}
\affiliation{Department of Physics, University of Virginia, Charlottesville, Virginia 22904, USA.}

\begin{abstract}
Despite the tremendous success of general relativity so far, modified theories of gravity have received increased attention lately, motivated from both theoretical and observational aspects.
Gravitational wave observations opened new possibilities for testing the viability of such theories in the strong-field regime.
One could test each theory against observed data one at a time, though a more efficient approach would be to first probe gravity in a theory-agnostic way and map such information to that on specific theories afterward.
One example of such model-independent tests with gravitational waves is the parameterized post-Einsteinian formalism, where one introduces generic parameters in the amplitude and phase that capture non-Einsteinian effects.  
In this paper, we derive gravitational waveforms from inspiraling compact binaries in various modified theories of gravity that violate at least one fundamental pillar in general relativity.
We achieve this by deriving relations between corrections to the waveform amplitude/phase and those to the frequency evolution and Kepler's third law, since the latter two have already been (or can easily be) derived in many theories.
Such an analysis allows us to derive corrections to the waveform amplitude, which extends many of previous works that focused on phase corrections only.
Moreover, we derive gravitational waveforms in theories with varying gravitational constant. We extend the previous work by introducing two gravitational constants (the conservative one entering in the binding energy and the dissipative one entering in the gravitational wave luminosity) and allowing masses of binary constituents to vary with time.
We also correct some errors in previous literature.  
Our results can be used to improve current analyses of testing general relativity as well as to achieve new projected constraints on many modified theories of gravity.

\end{abstract}

\date{\today}

\maketitle


\section{Introduction}
General relativity (GR) is one of the cornerstones of modern physics, and so far the most  successful theory of gravitation. Along with the elegant mathematical structure and solid conceptual foundation, GR has passed all the tests with high accuracy~\cite{Will:2014kxa}. However, there are theoretical and observational motivations which lead to the demand of a modified theory of gravitation. Regarding the former, GR is a purely classical theory and incompatible with quantum mechanics. Strong gravitational fields at Planck scale where quantum effects cannot be ignored~\cite{Adler:2010wf,Ng:2003jk}, such as in the vicinity of black holes (BHs) and the very early universe, require a consistent theory of quantum gravity for their complete description. Regarding the latter, puzzling observations such as the accelerated expansion of the universe~\cite{Abbott:1988nx,Copeland:2006wr,Perlmutter:1998np,Riess:1998cb,Riess:2004nr,RevModPhys.61.1,vanAlbada:1984js,WEINBERG201387} and anomalous kinematics of galaxies~\cite{article,Bosma:1981zz,Begeman:1991iy,Rubin:1970zza,Rubin:1980zd,1973ApJ...186..467O,Ostriker:1993fr} also suggest that one may need to go beyond GR to explain such cosmological phenomena if one does not wish to introduce dark energy or dark matter that are currently unknown.

Before gravitational waves (GWs) were directly detected by Advanced LIGO and Virgo, tests of gravity mainly focused on using solar system experiments and observations of radio pulsars and cosmology. Each of these cover different ranges of length scale and curvature strength. Solar system experiments constrain gravity in the weak-field and slow-motion environment. In terms of relativistic equations of motion, such experiments give access mostly to first order corrections to Newtonian dynamics~\cite{Berti:2015itd,Will:2014kxa}. Pulsar timing observations of neutron stars (NSs) offer us both weak-field and strong-field tests of gravity~\cite{Freire:2012mg,Kramer:2006nb,Liu:2011ae,Ransom:2014xla,Stairs:2002cw,Stairs:1997kz,Stairs:2003eg,taylor1992experimental,Wex:2014nva,Yunes:2013dva}. 
On one hand, binary components are widely separated and the relative motion of two stars in a binary is slow (and thus weak-field). On the other hand, binary pulsars consist of NSs which are compact and are strong-field sources of gravity.
Cosmological observations constrain gravity in the weak field regime but at length scales which are many orders of magnitude larger compared to other tests~\cite{Berti:2015itd,Clifton:2011jh,Jain:2010ka,Joyce:2014kja,Koyama:2015vza}. Cosmological tests of gravity include observations of cosmic microwave background radiation~\cite{0004-637X-737-2-98,Ade:2015xua,Salvatelli:2016mgy,Ade:2015xua,Bennett:2012zja,Hinshaw:2012aka}, studies of Big Bang Nucleosynthesis~\cite{Clifton:2005xr,Coc:2006rt,Damour:1998ae,Komatsu:2010fb,Mathews:2017xht,Olive:1999ij,PhysRevD.56.7627}, weak gravitational lensing~\cite{Bartelmann:1999yn,Collett:2018gpf,Clowe:2006eq,Huterer2010,Lewis:2006fu} and observations of galaxies~\cite{Berti:2015itd}. 
Other tests include using the orbital motion of stars near the Galactic Center~\cite{Ghez:2003qj,Hees:2017aal,Abuter:2018drb}. 

Up until now, six GW sources have been discovered (five of them being consistent with binary BH mergers~\cite{Abbott:2016blz,Abbott:2016nmj,Abbott:2017vtc,Abbott:2017gyy,Abbott:2017oio} while the remaining one being consistent with a binary NS merger~\cite{TheLIGOScientific:2017qsa}), which opened completely new ways of testing GR. 
GWs provide the opportunity to probe gravity in the strong-field and highly dynamical regime.
Binary BH merger events have been used to carry out a model-independent test of gravity by estimating the amount of residuals in the detected signals of GW150914 from the best-fit waveform~\cite{TheLIGOScientific:2016src}. GW150914 has also been used to perform a consistency test of GR between the inspiral and post-inspiral phases~\cite{TheLIGOScientific:2016src}. An addition of Virgo allowed one to look for non-tensorial polarization modes of GWs~\cite{Abbott:2017oio}. Meanwhile, the arrival time difference between gravitons and photons in the binary NS merger event GW170817 can be used to constrain the deviation in the propagation speed of the former from the latter to one part in $10^{15}$, to place bounds on the violation of Lorentz invariance and to carry out a new test of the equivalence principle via the Shapiro time delay~\cite{Monitor:2017mdv}.
Such a constraint on the propagation speed of GWs has led one to rule out many of modified theories of gravity that can explain the current accelerating expansion of our universe without introducing dark energy~\cite{Baker:2017hug,Creminelli:2017sry,Ezquiaga:2017ekz,Battye:2018ssx,Ezquiaga:2018btd,Sakstein:2017xjx,Lombriser:2016yzn,Lombriser:2015sxa}.
So far, no evidence has been found that indicates non-GR effects.

One can carry out yet another type of tests of GR by directly measuring or constraining non-GR parameters in the waveform. One can derive modifications to GR waveforms by choosing specific modified theories of gravity, though perhaps a more efficient approach is to perform the test in a model-independent way. A pioneering work along this line has been carried out in~\cite{Arun:2006yw,Arun:2006hn,Mishra:2010tp}, where the authors treat each post-Newtonian (PN) term in the waveform independently and look for consistency among them. Based on this, a data analysis pipeline (TIGER) was developed~\cite{Agathos:2013upa,Meidam:2014jpa}. One drawback of such a formalism is that one can only treat PN terms in non-GR theories that are also present in GR, which means that one cannot capture e.g. scalar dipole radiation effect entering at a negative PN order that is absent in GR. To overcome this, Yunes and Pretorius~\cite{Yunes:2009ke} proposed a new framework called \emph{parameterized post-Einsteinian} (PPE) formalism, where they introduced new parameters that can capture non-GR effects in waveforms in a generic way. The original work focused on tensorial polarizations for quasi-circular binaries and introduced only the leading PN non-GR corrections in Fourier domain. Such an analysis was later extended to include non-tensorial polarizations~\cite{Chatziioannou:2012rf} and multiple PN correction terms~\cite{Sampson:2013lpa}, and for time domain waveforms~\cite{Huwyler:2014gaa}, eccentric binaries~\cite{Loutrel:2014vja} and a sudden turn on of non-GR effects~\cite{Sampson:2013jpa,Sampson:2014qqa}. The LIGO Scientific Collaboration and Virgo Collaboration developed a generalized IMRPhenom model~\cite{Monitor:2017mdv} that is similar to the PPE formalism~\cite{Yunes:2016jcc}. Generic non-GR parameters in the waveform phase have been constrained in~\cite{Monitor:2017mdv,TheLIGOScientific:2016pea,Yunes:2016jcc,Abbott:2017vtc} with the observed GW events.

In this paper, we derive PPE waveforms in various modified theories of gravity. Many of previous literature focused on deriving phase corrections since matched filtering is more sensitive to such phase corrections than to amplitude corrections. Having said this, there are situations where amplitude corrections are more useful to probe, such as amplitude birefringence in parity-violating theories of gravity~\cite{Alexander:2007kv,Yunes:2008bu,Yunes:2010yf,Yagi:2017zhb} and testing GR with astrophysical stochastic GW backgrounds~\cite{Maselli:2016ekw}. We first derive PPE amplitude and phase corrections in terms of generic modifications to the frequency evolution and Kepler's third law that determine the waveform in Fourier domain. For our purpose, this formalism is more useful than that in~\cite{Chatziioannou:2012rf}, which derives the amplitude and phase corrections in terms of generic modifications to the binding energy of a binary and the GW luminosity. 
We follow the original PPE framework and focus on deriving leading PN corrections in tensorial modes only~\cite{Yunes:2009ke,Cornish:2011ys}. Non-tensorial GW modes also typically exist in theories beyond GR, though at least in scalar-tensor theories, the amplitude of a scalar polarization is of higher PN order than amplitude corrections to tensor modes~\cite{Chatziioannou:2012rf,Arun:2012hf}.

We also derive non-GR corrections in varying-$G$ theories, considering a PPE formalism with variable gravitational constants. Although there is only one gravitational constant in GR, many modified theories allow more than one gravitational constants that appear in different sectors. We consider two different gravitational constants, one entering in the GW luminosity and the other in Kepler's third law or the binding energy. We also promote the binary  masses and the specific angular momentum to vary with time via the sensitivities~\cite{PhysRevLett.65.953}, which closely follow testing variation in $G$ with binary pulsars~\cite{Wex:2014nva}. Our work extends the previous work of Ref.~\cite{Yunes:2009bv} where dissipative and conservative constants were taken to be the same and the masses of binary components were assumed to be constant. Furthermore, we correct the energy-balance law used in~\cite{Yunes:2009bv} for varying-$G$ theories by taking into account the non-conservation of binding energy in the absence of gravitational radiation.

Non-GR corrections can enter in the gravitational waveform through activation of different theoretical mechanisms, which can be classified as generation mechanisms and propagation mechanisms~\cite{Yunes:2016jcc}. Generation mechanisms take place close to the source (binary), while propagation mechanisms occur in the far-zone and accumulate over distance as the waves propagate. In this paper, we focus on the former\footnote{PPE waveforms due to modifications in the propagation sector can be found in~\cite{Mirshekari:2011yq,Yunes:2016jcc,Nishizawa:2017nef}, which have been used for GW150914, GW151226~\cite{Yunes:2016jcc} and GW170104~\cite{Abbott:2017vtc} to constrain the mass of the graviton and Lorentz violation.}. The PPE parameters in various modified theories of gravity are summarized in Tables~\ref{table:ppE_phase} (phase corrections) and~\ref{table:ppE_amp} (amplitude corrections).
Some of the amplitude corrections were derived here for the first time. We also correct some errors in previous literature.

The rest of the paper is organized as follows: In Sec.~\ref{section:ppE}, we revisit the standard PPE formalism. In Sec.~\ref{section:Example}, we derive the PPE parameters in some example theories following the formalism in Sec.~\ref{section:ppE}. In Sec.~\ref{gdot}, we derive the PPE parameters in varying-$G$ theories. We summarize our work and discuss possible future prospects in Sec.~\ref{conclusions}. Appendix~\ref{appendix} discusses the original PPE formalism. In App.~\ref{appendix_2}, we derive the frequency evolution in varying-$G$ theories from the energy-balance law. We use the geometric units $G=c=1$ throughout this paper except for varying-$G$ theories. 

{
\newcommand{\minitab}[2][l]{\begin{tabular}{#1}#2\end{tabular}}
\renewcommand{\arraystretch}{2.}
\begingroup 
\begin{table*}[htb]
\begin{centering}
\begin{tabular}{c|c|c|c}
\hline
\hline
\noalign{\smallskip}
\multirow{2}{*}{Theories}&\multicolumn{2}{c|}{PPE Phase Parameters} & \multirow{2}{*}{Binary Type}\\ \cline{2-3}
& Magnitude ($\beta$)   & Exp. ($b$) &  \\ \hline
 Scalar-Tensor~\cite{Scharre:2001hn,Berti:2004bd}&$-\frac{5}{7168}\eta ^{2/5}(\alpha_1-\alpha_2)^2$& $-7$ & Any\\ \hline
EdGB~\cite{Yagi:2011xp}&$-\frac{5}{7168}\zeta_\EDGB\frac{\left(m_1^2\tilde s_2^\EDGB-m_2^2\tilde s_1^\EDGB\right)^2}{m^4\eta^{18/5}}$&$-7$ & Any\\ \hline
DCS~\cite{Yagi:2012vf,Yunes:2016jcc}&$\frac{481525}{3670016}\eta ^{-14/5} \zeta_{\DCS} \left[-2 \delta_m \chi_a \chi_s+\left(1-\frac{4992 \eta }{19261}\right)\chi_a^2 +\left(1-\frac{72052 \eta }{19261}\right)\chi_s^2\right]
$&$-1$ & BH/BH\\ \hline
Einstein-\AE ther~\cite{Hansen:2014ewa}& $-\frac{5}{3584}\eta ^{2/5} \frac{(s_1^\EA-s_2^\EA)^2}{[(1-s_1^\EA) (1-s_2^\EA)]^{4/3}}\left[ \frac{(c_{14}-2) w_0^3-w_1^3}{c_{14} w_0^3 w_1^3}\right]$  & $-7$ & Any\\ \hline
Khronometric~\cite{Hansen:2014ewa}&$-\frac{5}{3584} \eta ^{2/5}\frac{(s_1^\KG-s_2^\KG)^2}{[(1-s_1^\KG) (1-s_2^\KG)]^{4/3}}  \sqrt{\bar \alpha_{\KG}}\left[\frac{(\bar \beta_{\KG}-1)(2+\bar \beta_{\KG}+3\bar \lambda_{\KG})}{(\bar \alpha_{\KG}-2)(\bar \beta_{\KG}+\bar \lambda_{\KG})}\right]^{3/2}$&$-7$ & Any\\ \hline 
Noncommutative~\cite{Kobakhidze:2016cqh}& $-\frac{75}{256}\eta ^{-4/5}(2 \eta -1) \Lambda ^2$ &$-1$ &  BH/BH\\ \hline
Varying-$G$~\cite{Yunes:2009bv}&$\mathbf{-\frac{25}{851968}\eta_0^{3/5}\dot{G}_{C,0}\left[11m_0 +3(s_{1,0}+s_{2,0}-\delta_\Gdot)m_0-41 (m_{1,0}s_{1,0}+m_{2,0}s_{2,0})\right]}$&$ \mathbf{-13}$ & Any\\ 
\noalign{\smallskip}
\hline
\hline
\end{tabular}
\end{centering}
\caption{PPE corrections to the GW phase $\delta\Psi\equiv\beta u^b$ in Fourier space in various modified theories of gravity, where $\beta$ is the magnitude correction (second column) and $b$ is the exponent correction (third column). $u\equiv(\pi G_C \mathcal{M}f)^{1/3}$, where $\mathcal{M}$ and $\eta$ are the chirp mass and the symmetric mass ratio of the binary respectively, and $G_C$ is the conservative gravitational constant appearing in Kepler's third law. We adopt the unit $G_C\equiv 1$ in all theories except for the varying-$G$ ones. 
The expressions in dynamical Chern-Simons (dCS) gravity and noncommutative gravity only apply to binary BHs, while those in other theories apply to any compact binaries (last column)\footnote{Practically speaking, if NSs are spinning much slower than BHs, one can use the dCS expression also for BH/NS binaries by setting one of the spins to zero.}.
The mass, sensitivity, and scalar charge of the $A$th binary component are represented by $m_A$, $s_A$, and $\alpha_A$ respectively. 
$\zeta_\EDGB$ and $\zeta_{\DCS}$ are the dimensionless coupling constants in Einstein-dilaton Gauss-Bonnet (EdGB) and dCS gravity respectively. $\tilde s_{A}^\EDGB$ are the spin-dependent factors of the scalar charges in EdGB gravity, given below Eq.~\eqref{eq:beta-EdGB} for BHs while 0 for ordinary stars. 
$\chi_{s,a}$ are the symmetric and antisymmetric combinations of dimensionless spin parameters and $\delta_m$ is the fractional difference in masses relative to the total mass $m$. The amount of Lorentz violation in Einstein-\AE ther theory and khronometric gravity is controlled by $(c_1,c_2,c_3,c_4)$ and $(\bar \alpha_{\KG}, \bar \beta_{\KG}, \bar \lambda_{\KG})$ respectively. $w_s$ is the propagation speed of the spin-$s$ modes in Einstein-\AE ther theory given by Eqs.~\eqref{eq:Prop_Speed_EA_1}-\eqref{eq:Prop_Speed_EA_3}, and $c_{14}\equiv c_1+c_4$. The representative parameter in noncommutative gravity is $\Lambda$. 
The subscript $0$ in varying-$G$ theories denotes that the quantity is measured at the time of coalescence $t_0$, while a dot refers to a time derivative.  
$\delta_\Gdot$ is the fractional difference between the rates at which conservative and dissipative  gravitational constants change in time. The former is $G_C$ as already explained while the dissipative gravitational constant is defined as the one that enters in the GW luminosity through Eq.~\eqref{eq:2h2}.
The boldface expression indicates that it has been derived here for the first time.
}
\label{table:ppE_phase}
\end{table*}
\endgroup
}

{
\newcommand{\minitab}[2][l]{\begin{tabular}{#1}#2\end{tabular}}
\renewcommand{\arraystretch}{2.}
\begingroup 
\begin{table*}[htb]
\begin{centering}
\begin{tabular}{c|c|c}
\hline
\hline
\noalign{\smallskip}
\multirow{2}{*}{Theories}&\multicolumn{2}{c}{PPE Amplitude Parameters}\\ \cline{2-3}
& Magnitude ($\alpha$)& Exponent ($a$)  \\ \hline
Scalar-Tensor~\cite{Arun:2012hf,Chatziioannou:2012rf,Liu:2018sia} &$-\frac{5}{192}\eta ^{2/5}(\alpha_1-\alpha_2)^2$&$-2$\\ \hline
EdGB &$\mathbf{-\frac{5}{192}\zeta_\EDGB\frac{\left(m_1^2 \tilde s_2^\EDGB-m_2^2 \tilde s_1^\EDGB \right)^2}{m^4\eta^{18/5}}}$&$\mathbf{-2}$\\ \hline
DCS &$\mathbf{\frac{57713}{344064}\eta ^{-14/5}\zeta_{\DCS} \left[-2 \delta_m \chi_a \chi_s+\left(1-\frac{14976 \eta }{57713}\right) \chi_a^2+\left(1-\frac{215876 \eta }{57713}\right) \chi_s^2\right]}
$&$\mathbf{+4}$\\ \hline
Einstein-\AE ther~\cite{Hansen:2014ewa}& $\mathbf{-\frac{5}{96}\eta ^{2/5} \frac{(s_1^\EA-s_2^\EA)^2}{[ (1-s_1^\EA) (1-s_2^\EA)]^{4/3}}\left[\frac{(c_{14}-2) w_0^3-w_1^3}{c_{14} w_0^3 w_1^3}\right]}$& $\mathbf{-2}$\\ \hline
Khronometric~\cite{Hansen:2014ewa}&$\mathbf{-\frac{5}{96} \eta ^{2/5}\frac{(s_1^\KG-s_2^\KG)^2}{[(1-s_1^\KG) (1-s_2^\KG)]^{4/3}}  \sqrt{\bar \alpha_{\KG}}\left[\frac{(\bar \beta_{\KG}-1)(2+\bar \beta_{\KG}+3\bar \lambda_{\KG})}{(\bar \alpha_{\KG}-2)(\bar \beta_{\KG}+\bar \lambda_{\KG})}\right]^{3/2}}$&$\mathbf{-2}$ \\ \hline 
Noncommutative & $\mathbf{-\frac{3}{8}\eta ^{-4/5}(2 \eta -1) \Lambda ^2}$ &$\mathbf{+4}$\\ \hline
Varying-$G$~\cite{Yunes:2009bv}& $\mathbf{\frac{5}{512}\eta_0^{3/5}\dot{G}_{C,0}\left[-7m_0+(s_{1,0}+s_{2,0}-\delta_\Gdot)m_0+13 (m_{1,0}s_{1,0}+m_{2,0}s_{2,0})\right]}$ &$\mathbf{-8}$\\ 
\noalign{\smallskip}
\hline
\hline
\end{tabular}
\end{centering}
\caption{PPE corrections to the GW amplitude $|\tilde h| = |\tilde{h}_{\GR}| (1+\alpha u^a)$ in Fourier space in various modified theories of gravity with the magnitude $\alpha$ (second column) and the exponent $a$ (third column), and $|\tilde{h}_{\GR}|$ representing the amplitude in GR. The meaning of other parameters are the same as in Table~\ref{table:ppE_phase}. The expressions in boldface correspond to either those derived here for the first time or corrected expressions from previous literature. 
}
\label{table:ppE_amp}
\end{table*}
\endgroup
}

\section{PPE Waveform}\label{section:ppE}
We begin by reviewing the PPE formalism. The original formalism (that we explain in detail in App.~\ref{appendix}) was developed by considering non-GR corrections to the binding energy $E$ and GW luminosity $\dot E$~\cite{Yunes:2009ke,Chatziioannou:2012rf}. The former (latter) correspond to conservative (dissipative) corrections. Here, we take a slightly different approach and consider corrections to the GW frequency evolution $\dot f$ and the Kepler's law $r(f)$, where $r$ is the orbital separation while $f$ is the GW frequency. This is because these two quantities directly determine the amplitude and phase corrections away from GR, and hence, the final expressions are simpler than the original ones. Moreover, non-GR corrections to $\dot f$ and $r(f)$ have already been derived in previous literature for many modified theories of gravity.

PPE gravitational waveform for a compact binary inspiral in Fourier domain is given by~\cite{Yunes:2009ke}
\begin{equation}\label{eq:2a}
\tilde{h}(f)=\tilde{h}_{\GR}(1+\alpha\, u^a)e^{i\delta\Psi}\,,
\end{equation}
where $\tilde{h}_{\GR}$ is the gravitational waveform in GR. $\alpha\, u^a$ corresponds to the non-GR correction to the GW amplitude while  $\delta \Psi$ is that to the GW phase with
\begin{equation}
u=(\pi \mathcal{M} f)^\frac{1}{3}\,.
\end{equation}
$\mathcal{M}=(m_1m_2)^{3/5}/(m_1+m_2)^{1/5}$ is the chirp mass with component masses $m_1$ and $m_2$. $u$ is proportional to the relative velocity of the binary components. $\alpha$ represents the overall magnitude of the amplitude correction while $a$ gives the velocity dependence of the correction term. In a similar manner, one can rewrite the phase correction as 
\begin{equation}\label{eq:2b}
\delta\Psi=\beta u^b\,.
\end{equation}
$\alpha$, $\beta$, $a$, and $b$ are called the PPE parameters. When $(\alpha,\beta)\equiv(0,0)$, Eq.~\eqref{eq:2a} reduces to the waveform in GR.

One can count the PN order of non-GR corrections in the waveform as follows. A correction term is said to be of $n$~PN relative to GR if the \emph{relative} correction is proportional $u^{2n}$. Thus, the amplitude correction in Eq.~\eqref{eq:2a} is of $a/2$~PN order. On the other hand, given that the leading GR phase is proportional to $u^{-5}$ (see Eq.~\eqref{eq:Psi_GR}), the phase correction in Eq.~\eqref{eq:2b} is of $(b+5)/2$~PN order.

As we mentioned earlier, the PPE modifications in Eq.~\eqref{eq:2a} enter through corrections to the orbital separation and the frequency evolution.
We parameterize the former as
 \begin{equation}
 \label{eq:2k}
 r=r_{\GR}(1+\gamma_r u^{c_r})\,,
 \end{equation}
where $\gamma_r$ and $c_r$ are non-GR parameters which show the deviation of the orbital separation $r$ away from the GR contribution $r_{\GR}$. To leading PN order, $r_{\GR}$ is simply given by the Newtonian Kepler's law as $r_{\GR}=\left(m/\Omega^2\right)^{1/3}$. Here $m\equiv m_1+m_2$ is the total mass of the binary while $\Omega\equiv\pi f $ is the orbital angular frequency. The above correction to the orbital separation arises purely from conservative corrections (namely corrections to the binding energy).

Similarly, we parameterize the GW frequency evolution with non-GR parameters $\gamma_{\dot{f}}$ and $c_{\dot{f}}$  as
\begin{equation}\label{eq:2m}
\dot{f}=\dot{f}_{\GR}\left(1+\gamma_{\dot{f}}u^{c_{\dot{f}}}\right)\,.
\end{equation}
Here $\dot{f}_{\GR}$ is the frequency evolution in GR which, to leading PN order, is given by~\cite{cutlerflanagan,Blanchet:1995ez}
\begin{align}\label{eq:2s}
\dot{f}_{\GR}=\frac{96}{5}\pi^{8/3}\mathcal{M}^{5/3}f^{11/3}=\frac{96}{5\pi\mathcal{M}^2}u^{11}\,.
\end{align}
Unlike the correction to the orbital separation, the one to the frequency evolution originates corrections from both the conservative and dissipative sectors.

Below, we will derive how the PPE parameters $(\alpha, \beta, a, b)$ are given in terms of $(\gamma_r,c_r)$ and $(\gamma_{\dot{f}},c_{\dot{f}})$. We will also show how the amplitude PPE parameters $(\alpha, a)$ can be related to the phase PPE ones $(\beta,b)$ in certain cases. We will assume that non-GR corrections are always smaller than the GR contribution and keep only to leading order in such corrections at the leading PN order.

 \subsection{Amplitude Corrections}
 
Let us first look at corrections to the waveform amplitude.  Within the stationary phase approximation~\cite{PhysRevD.62.084036,Yunes:2009yz}, the waveform amplitude for the dominant quadrupolar radiation in Fourier domain is given by
 \begin{equation}\label{eq:2h1}
\tilde{\mathcal{A}}(f)=\frac{A(\bar{t})}{2\sqrt{\dot{f}}}\,.
\end{equation}
Here $A$ is the waveform amplitude in the time domain while $\bar t (f)$ represents time at the stationary point. $\mathcal{A}(\bar t)$ can be obtained by using the quadrupole formula for the metric perturbation in the transverse-traceless gauge given by~\cite{Blanchet:2002av}
 \begin{equation}\label{eq:2h2}
h^{ij}(t)\propto \frac{G}{D_L}\frac{d^2 }{d t^2}Q^{ij}\,.
 \end{equation}
Here $D_L$ is the source's luminosity distance and $Q^{ij}$ is the source's quadruple moment tensor. 

For a quasi-circular compact binary, $\tilde{\mathcal{A}}$ in Eq.~\eqref{eq:2h1} then becomes
\begin{equation}\label{eq:2e}
\tilde{\mathcal{A}}(f) \propto\frac{1}{\sqrt{\dot{f}}}\frac{G}{D_L}\mu r^2f^2 \propto \frac{r^2}{\sqrt{\dot{f}}}  \,,
\end{equation}
where $\mu$ is the reduced mass of the binary.
Substituting Eqs.~\eqref{eq:2k} and~\eqref{eq:2m} into Eq.~\eqref{eq:2e} and keeping only to leading order in non-GR corrections, we find
\begin{equation}\label{eq:2n}
\tilde{\mathcal{A}}(f)=\tilde{\mathcal{A}}_{\GR} \left(1+2\gamma_ru^{c_r}-\frac{1}{2}\gamma_{\dot{f}}u^{c_{\dot{f}}}\right)\,,
\end{equation}
where $\tilde{\mathcal{A}}_{\GR} $ is the amplitude of the Fourier waveform in GR. Notice that this expression is much simpler than that in the original formalism in Eq.~\eqref{eq:o2}.

Let us now show the expressions for the PPE parameters $\alpha$ and $a$ for three different cases using Eq.~\eqref{eq:2n}:

\begin{itemize}

\item
\emph{Dissipative-dominated Case}

When dissipative corrections dominate, we can neglect corrections to the binary separation $(\gamma_{r} = 0)$ and Eq.~\eqref{eq:2n} reduces to
\begin{equation}
\tilde{\mathcal{A}}(f)=\tilde{\mathcal{A}}_{\GR} \left(1-\frac{1}{2}\gamma_{\dot{f}}u^{c_{\dot{f}}}\right)\,.
\end{equation}
Comparing this with the PPE waveform in Eq.~\eqref{eq:2a}, we find
\begin{equation}\label{eq:2t}
\alpha=-\frac{\gamma_{\dot{f}}}{2},\quad a=c_{\dot{f}}\,.
\end{equation}

\item
\emph{Conservative-dominated Case}

When conservative corrections dominate, $c_r = c_{\dot f}$ and there is an explicit relation between $\gamma_{r}$ and $\gamma_{\dot f}$. Though finding such a relation is quite involved and one needs to go back to the original PPE formalism as explained in App.~\ref{appendix}. Non-GR corrections to the GW amplitude in such a formalism is shown in Eq.~\eqref{eq:o}. Setting the dissipative correction to zero, one finds 
\begin{align}\label{eq:2u}
\alpha=-\frac{\gamma_r}{a}(a^2-4a-6)\,, \quad a=c_r = c_{\dot f}\,.
\end{align} 

\item
\emph{Comparable Dissipative and Conservative Case}

If dissipative and conservative corrections enter at the same PN order, we can set $c_r=c_{\dot{f}}$ in Eq.~\eqref{eq:2n}. Since there is no generic relation between $\gamma_{r}$ and $\gamma_{\dot f}$ in this case, one simply finds  
\begin{equation}
\label{eq:amp-ppE-comparable}
\alpha=2 \text{$\gamma_r $}-\frac{\text{$\gamma_{\dot{f}} $}}{2}\,, \quad a=c_{r}=c_{\dot{f}}\,.
\end{equation}

\end{itemize}
Example modified theories of gravity that we study in Secs.~\ref{section:Example} and~\ref{gdot} fall into either the first or third case.

\subsection{Phase Corrections}

Next, let us study corrections to the GW phase. The phase $\Psi$ in Fourier domain is related to the frequency evolution as~\cite{Tichy:1999pv}
\begin{equation}
\frac{d^2\Psi}{d \Omega^2}=2\frac{d t}{d\Omega}\,,
\end{equation}
which can be rewritten as
\begin{equation}
\frac{d^2\Psi}{d \Omega^2}=\frac{2}{\pi \dot{f}}\,.
\end{equation}
Substituting Eq.~\eqref{eq:2m} to the right hand side of the above equation and keeping only to leading non-GR correction, we find 
\begin{equation}\label{eq:2q}
\frac{d^2\Psi}{d \Omega^2}=\frac{2}{\pi\dot{f}_{\GR}}(1-\gamma_{\dot{f}}u^{c_{\dot{f}}})\,.
\end{equation}
Using further Eq.~\eqref{eq:2s} to Eq.~\eqref{eq:2q} gives
\begin{equation}\label{eq:2r}
\frac{d^2\Psi}{d \Omega^2}=\frac{5}{48}\mathcal{M}^2u^{-11}(1-\gamma_{\dot{f}}u^{c_{\dot{f}}})\,.
\end{equation}

We are now ready to derive $\Psi$ and extract the PPE parameters $\beta$ and $b$. Using $\Omega = \pi f$, we can integrate Eq.~\eqref{eq:2r} twice to find
\be
\label{eq:Psi}
\Psi = \Psi_\GR  -\frac{15 \text{$\gamma_{\dot{f}} $}}{16 (\text{$c_{\dot{f}}$}-8) (\text{$c_{\dot{f}}$}-5)} u^{c_{\dot{f}}-5}
\ee
for $c_{\dot{f}} \neq 5$ and $c_{\dot{f}} \neq 8$. Here we only keep to leading non-GR correction and $\Psi_{\GR}$ is the GR contribution given in Eq.~\eqref{eq:Psi_GR} to leading PN order. Similar to the amplitude case, the above expression is much simpler than that in the original formalism in Eq.~\eqref{eq:p2}.
Comparing this with Eqs.~\eqref{eq:2a} and~\eqref{eq:2b}, we find
\begin{equation}
\label{eq:2v}
\beta=-\frac{15 \text{$\gamma_{\dot{f}} $}}{16 (\text{$c_{\dot{f}}$}-8) (\text{$c_{\dot{f}}$}-5)}\,, \quad b=c_{\dot{f}}-5\,.
\end{equation}
The above relation is valid for all three types of corrections considered for the GW amplitude case. 

In App.~\ref{appendix}, we review $\delta\Psi$ derived in the original PPE formalism, where we show dissipative and conservative contributions explicitly. In particular, one can use Eq.~\eqref{eq:p} to find $\beta$ for all three cases separately.

\subsection{Relations among ppE Parameters}

Finally, we study relations among the PPE parameters. From Eqs.~\eqref{eq:2t}--\eqref{eq:amp-ppE-comparable} and~\eqref{eq:2v}, one can easily see 
\begin{equation}
b=a-5\,,
\end{equation}
which holds in all three cases considered previously. Let us consider such three cases in turn below to derive relations between $\alpha$ and $\beta$. 

\begin{itemize}

\item
\emph{Dissipative-dominated Case}

When dissipative corrections dominate, we can use Eqs.~\eqref{eq:2t} and~\eqref{eq:2v} to find $\alpha$ in terms of $\beta$ and $a$ as
\begin{equation}\label{eq:2w2}
\alpha = \frac{8}{15} (a-8)(a-5) \, \beta\,.
\end{equation}

\item
\emph{Conservative-dominated Case}

When conservative corrections dominate, we can set the dissipative correction to vanish in Eq.~\eqref{eq:p} to find 
\begin{equation}\label{eq:2w}
\beta=-\frac{15}{8}\frac{\gamma_r}{c_r}\frac{c_r^2-2c_r-6}{(8-c_r)(5-c_r)}\,, \quad b=c_r-5\,.
\end{equation}
Using this equation together with Eq.~\eqref{eq:2u}, we find
\begin{equation}
\alpha =\frac{8}{15} \frac{(8-a)(5-a)(a^2-4a-6)}{a^2-2a-6} \beta\,.
\end{equation}

\item
\emph{Comparable Dissipative and Conservative Case}

When dissipative and conservative corrections enter at the same PN order, there is no explicit relation between $\alpha$ and $\beta$. This is because $\alpha$ depends both on $\gamma_r$ and $\gamma_{\dot f}$ (see Eq.~\eqref{eq:amp-ppE-comparable}) while $\beta$ depends only on the latter (see Eq.~\eqref{eq:Psi}), and there is no relation between the former and the latter. Thus, one can rewrite $\gamma_{\dot f}$ in terms of $\beta$ and substitute into Eq.~\eqref{eq:amp-ppE-comparable} but cannot eliminate $\gamma_r$ from the expression for $\alpha$. 

\end{itemize}

 \section{Example Theories}\label{section:Example}
In this section, we consider several modified theories of gravity where non-GR corrections arise from generation mechanisms. We briefly discuss each theory, describing differences from GR and its importance. We derive the PPE parameters for each theory following the formalism in Sec.~\ref{section:ppE}. Among the various example theories we present here, dissipative corrections dominate in scalar-tensor theories, EdGB gravity, Einstein-\AE ther theory, and khronometric gravity. On the other hand, dissipative and conservative corrections enter at the same PN order in dCS gravity, noncommutative gravity, and varying-$G$ theories. We do not consider any theories where conservative corrections dominate dissipative ones, though such a situation can be realized for e.g. equal-mass and equal-spin binaries in dCS gravity, where the scalar quadrupolar radiation is suppressed and dominant corrections arise from the scalar dipole interaction and quadrupole moment corrections in the conservative sector.

 \subsection{Scalar-Tensor Theories}
Scalar-tensor theories are one of the most well-established modified theories of gravity where at least one scalar field is introduced through a non-minimal coupling to gravity~\cite{Berti:2015itd,Chiba:1997ms,PhysRevD.6.2077}. Such theories arise naturally from the dimensional reduction of higher dimensional theories, such Kaluza-Klein theory~\cite{Fujii:2003pa,Overduin:1998pn} and string theories~\cite{polchinski1,polchinski2}.
Scalar-tensor theories have implications to cosmology as well since they are viable candidates for accelerating expansion of our universe~\cite{Brax:2004qh,PhysRevD.73.083510,PhysRevD.62.123510,PhysRevD.66.023525,Schimd:2004nq}, structure formation~\cite{Brax:2005ew}, inflation~\cite{Burd:1991ns,Barrow:1990nv,Clifton:2011jh}, and primordial nucleosynthesis~\cite{Coc:2006rt,Damour:1998ae,Larena:2005tu,Torres:1995je}. Such theories also offer simple ways to self-consistently model possible variations in Newton's constant~\cite{Clifton:2011jh} (as we discuss in Sec.~\ref{gdot}). One of the simplest scalar-tensor theories is Brans-Dicke (BD) theory, where a non-canonical scalar field is non-minimally coupled to the metric with an effective strength inversely proportional to the coupling parameter $\omega_{\BD}$~\cite{PhysRev.124.925,Scharre:2001hn}. So far the most stringent bound on the theory has been placed by the Cassini-Huygens satellite mission via Shapiro time delay measurement, which gives $\omega_{\BD}>4\times10^4$~\cite{Bertotti:2003rm}. Another class of scalar-tensor theories that has been studied extensively is Damour-Esposito-Far\`ese (DEF) gravity (or sometimes called quasi Brans-Dicke theory), which has two coupling constants $(\alpha_0,\beta_0)$. This theory reduces to BD theory when $\beta_0$ is set to 0 and $\alpha_0$ is directly related to $\omega_\BD$. This theory predicts nonperturbative spontaneous or dynamical scalarization phenomena for NSs~\cite{PhysRevLett.70.2220,Barausse:2012da}. 

When scalarized NSs form compact binaries, these systems emit scalar dipole radiation that changes the orbital evolution from that in GR. Such an effect can be used to place bounds on scalar-tensor theories. For example, combining observational orbital decay results from multiple binary pulsars, the strongest upper bound on $\beta_0$ that controls the magnitude of scalarization in DEF gravity has been obtained as $\beta_0\gtrsim -4.38$ at $90\%$ confidence level~\cite{Shao:2017gwu}. More recently, observations of a hierarchical stellar triple system PSR J0337+1715 placed strong bounds on the Strong Equivalence Principle (SEP) violation parameter\footnote{SEP violation parameter is defined as $\Delta=m_G/m_I-1$, where $m_G$ and $m_I$ are respectively the gravitational and inertial mass of a pulsar~\cite{Archibald:2018oxs}.} as $|\Delta|\lesssim 2\times 10^{-6}$ at $95\%$ confidence level~\cite{Archibald:2018oxs}. This bound stringently constrained the parameter space $(\alpha_0,\beta_0)$ of DEF gravity~\cite{PhysRevLett.70.2220,1970ApJ...161.1059N,Bergmann1968,Horbatsch:2010hj,PhysRevD.1.3209}.

Can BHs also possess scalar hair like NSs in scalar-tensor theories? BH no-hair theorem can be applied to many of scalar-tensor theories that prevents BHs to acquire scalar charges~\cite{Hawking:1972qk,Bekenstein:1995un,Sotiriou:2011dz,Hui:2012qt,Maselli:2015yva} including BD and DEF gravity, though exceptions exist, such as EdGB gravity~\cite{Yunes:2011we,Sotiriou:2013qea,Sotiriou:2014pfa,Silva:2017uqg,Doneva:2017bvd} that we explain in more detail in the next subsection. On the other hand, if the scalar field cosmologically evolves as a function of time, BHs can acquire scalar charges, known as the BH miracle hair growth~\cite{Jacobson:1999vr,Horbatsch:2011ye} (see also~\cite{Healy:2011ef,Berti:2013gfa} for related works). 

Let us now derive the PPE parameters in scalar tensor theories. Gravitational waveforms are modified from that in GR through the scalar dipole radiation. Using the orbital decay rate of compact binaries in scalar-tensor theories in~\cite{Freire:2012mg,Wex:2014nva}, one can read off the non-GR corrections to $\dot f$ as
\be
\gamma_{\dot f} = \frac{5}{96} \eta ^{2/5}(\alpha_1-\alpha_2)^2
\ee
with $c_{\dot f} = -2$.
Given that the leading correction to the waveform is the dissipative one in scalar-tensor theories, one can use Eq.~\eqref{eq:2v} to derive the PPE phase correction as
\be\label{eq:betaST}
\beta_{\ST}=-\frac{5}{7168}\eta ^{2/5}(\alpha_1-\alpha_2)^2
\ee
with $b=-7$. Here $\alpha_A$ represents the scalar charge of the $A$th binary component.
Using further Eq.~\eqref{eq:2w2}, one finds the amplitude correction as
\be\label{eq:alphaST}
\alpha_{\ST}=-\frac{5}{192}\eta ^{2/5}(\alpha_1-\alpha_2)^2
\ee
with $a=-2$. These corrections enter at $-1$~PN order relative to GR.

The scalar charges $\alpha_A$ depend on specific theories and compact objects. For example, in situations where the BH no-hair theorem~\cite{Hawking:1972qk,Bekenstein:1995un,Sotiriou:2011dz} applies, $\alpha_A = 0$. On the other hand, if the scalar field is evolving cosmologically, BHs undergo \emph{miracle hair growth}~\cite{Jacobson:1999vr} and acquire scalar charges given by~\cite{Horbatsch:2011ye}
\be
\alpha_A = 2 \, m_A \, \dot \phi\, [1+(1-\chi_A^2)^{1/2}]\,,
\ee
where $\dot{\phi}$ is the growth rate of the scalar field while $m_A$ and $\chi_A$ are the mass and the magnitude of the dimensionless spin angular momentum of the $\mathit{A}\text{th}$ body respectively. The PPE phase parameter $\beta$ for binary BHs in such a situation was derived in~\cite{Yunes:2016jcc}. Another well-studied example is Brans-Dicke theory, where one can replace $(\alpha_1-\alpha_2)^2$ in Eqs.~\eqref{eq:betaST} and~\eqref{eq:alphaST} as $2 (s_1-s_2)^2/(2+\omega_\BD)$~\cite{Freire:2012mg}. Here $s_A$ is the sensitivity of the $A$th body and roughly equals to its compactness (0.5 for BHs and $\sim 0.2$ for NSs). The PPE parameters in this theory has been found in~\cite{Chatziioannou:2012rf}. Scalar charges and the PPE parameters in generic screened modified gravity have recently been derived in~\cite{Zhang:2017srh,Liu:2018sia}.

The phase correction in Eq.~\eqref{eq:betaST} has been used to derive current and future projected bounds with GW interferometers. Regarding the former, GW150914 and GW151226 do not place any meaningful bounds on $\dot \phi$~\cite{Yunes:2016jcc}. On the other hand, by detecting GWs from BH-NS binaries, aLIGO and Virgo with their design sensitivities can place bounds that are stronger than the above binary pulsar bounds from dynamical scalarization for certain equations of state and NS mass range~\cite{Shibata:2013pra,Taniguchi:2014fqa,Sampson:2014qqa,Shao:2017gwu}\footnote{One needs to multiply Eq.~\eqref{eq:betaST} by a step-like function to capture the effect of dynamical scalarization.}. Einstein Telescope, a third generation ground-based detector, can yield constraints on BD theory from BH-NS binaries that are 100 times stronger than the current bound~\cite{Zhang:2017sym}. Projected bounds with future space-borne interferometers, such as DECIGO, can be as large as four orders of magnitude stronger than current bounds~\cite{Yagi:2009zz}, while those with LISA may not be as strong as the current bound~\cite{Berti:2004bd,Yagi:2009zm}.

Up until now, we have focused on theories with a massless scalar field, but let us end this subsection by commenting on how the above expressions for the PPE parameters change if one considers a massive scalar field instead. In such a case, the scalar dipole radiation is present only when the mass of the scalar field $m_s$ is smaller than the orbital angular frequency $\Omega= \pi f$. Then, if the Yukawa-type correction to the binding energy is subdominant, Eqs.~\eqref{eq:betaST} and~\eqref{eq:alphaST} simply acquire an additional factor of $\Theta(\Omega - m_s/\hbar)$, where $\Theta$ is the Heaviside function. For example, the gravitational waveform phase in massive BD theory is derived in~\cite{Berti:2012bp}. The situation is similar if massive pseudo-scalars are present, such as axions~\cite{Huang:2018pbu}.
 
  \subsection{Einstein-dilaton Gauss-Bonnet Gravity}
EdGB gravity is a well-known extension of GR, which emerges naturally in the framework of low-energy effective string theories  and gives one of the simplest viable high-energy modifications to GR~\cite{Moura:2006pz,Pani:2009wy}. It also arises as a special case of Horndeski gravity~\cite{Zhang:2017unx,Berti:2015itd}, which is the most generic scalar-tensor theory with at most second-order derivatives in the field equations. One obtains the EdGB action by adding a quadratic-curvature term to the Einstein-Hilbert action, where the scalar field (dilaton) is non-minimally coupled to the Gauss-Bonnet term with a coupling constant $\bar{\alpha}_\EDGB$~\cite{Kanti:1995vq}\footnote{We use barred quantities for coupling constants so that one can easily distinguish them from the PPE parameters.}. A stringent upper bound on such a coupling constant has been placed using the orbital decay measurement of a BH low-mass X-ray binary (LMXB) as $\sqrt{|\bar{\alpha}_\EDGB|} < 1.9\times10^5$ cm~\cite{Yagi:2012gp}. A similar upper bound has been placed from the existence of BHs~\cite{Pani:2009wy}. Equation-of-state-dependent bounds from the maximum mass of NSs have also been derived in~\cite{pani-EDGB-NS}.

BHs in EdGB gravity are of particular interest since they are fundamentally different from their GR counterparts. Perturbative but analytic solutions are available for static~\cite{Mignemi:1992nt,Mignemi:1993ce,Yunes:2011we,Sotiriou:2014pfa} and slowly rotating EdGB BHs~\cite{Pani:2011gy,Ayzenberg:2014aka,Maselli:2015tta} while numerical solutions have been found for static~\cite{Kanti:1995vq,Torii:1996yi,Alexeev:1996vs} and rotating~\cite{Pani:2009wy,Kleihaus:2011tg,Kleihaus:2014lba} BHs. One of the important reasons for considering BHs in EdGB is that BHs acquire scalar monopole charges~\cite{Yagi:2011xp,Sotiriou:2014pfa,Berti:2018cxi,Prabhu:2018aun} while ordinary stars such as NSs do not if the scalar field is coupled linearly to the Gauss-Bonnet term in the action~\cite{Yagi:2011xp,Yagi:2015oca}. This means that binary pulsars are inefficient to constrain the theory, and one needs systems such as BH-LMXBs~\cite{Yagi:2012gp} or BH/pulsar binaries~\cite{Yagi:2015oca} to have better probes on the theory.

We now show the expressions of the PPE parameters for EdGB gravity. The scalar monopole charge of EdGB BHs generates scalar dipole radiation, which leads to an earlier coalescence of BH binaries compared to GR. Such scalar radiation modifies the GW phase with the PPE parameters given by~\cite{Yunes:2016jcc,Yagi:2011xp}
\begin{equation}
\label{eq:beta-EdGB}
 \beta_\EDGB=-\frac{5}{7168}\zeta_\EDGB\frac{(m_1^2\tilde s_2^\EDGB-m_2^2\tilde s_1^\EDGB)^2}{m^4\eta^{18/5}}
 \end{equation}
 and $b=-7$.
 Here, $\zeta_\EDGB\equiv 16 \pi \bar{\alpha}_\EDGB^2/m^4$ is the dimensionless EdGB coupling parameter and $\tilde s_{A}^\EDGB$ are the spin-dependent factors of the BH scalar charges given by $\tilde s_{A}^\EDGB\equiv 2(\sqrt{1-{\chi_A}^2}-1+{\chi_A}^2)/{\chi_A}^2~$~\cite{Berti:2018cxi,Prabhu:2018aun}\footnote{$\tilde s_A^\EDGB$ are zero for ordinary stars like NSs~\cite{Yagi:2011xp,Yagi:2015oca}.}. In EdGB gravity, the leading order correction to the phase enters through the correction of the GW energy flux, and hence the theory corresponds to a dissipative-dominated case. We can then use Eq.~\eqref{eq:2w2} to calculate the amplitude PPE parameters as  
 \begin{equation}
 \alpha_\EDGB=-\frac{5}{192}\zeta_\EDGB\frac{(m_1^2 \tilde s_2^\EDGB-m_2^2 \tilde s_1^\EDGB)^2}{m^4\eta^{18/5}}
 \end{equation}
 and $a=-2$. These corrections enter at $-1$~PN order.
 
 One can use the phase correction in Eq.~\eqref{eq:beta-EdGB} to derive bounds on EdGB gravity with current~\cite{Yunes:2016jcc} and future~\cite{Yagi:2012gp} GW observations. Similar to the scalar-tensor theory case, current binary BH GW events do not allow us to place any meaningful bounds on the theory. Future second- and third-generation ground-based detectors and LISA can place bounds that are comparable to current bounds from LMXBs~\cite{Yagi:2012gp}. On the other hand, DECIGO has the potential to go beyond the current bounds by three orders of magnitude.
 
  \subsection{Dynamical Chern-Simons Gravity}
  
DCS gravity is described by Einstein-Hilbert action with a dynamical (pseudo-)scalar field which is non-minimally coupled to the Pontryagin density with a coupling constant $\bar{\alpha}_{\DCS}$~\cite{Alexander:2009tp,Jackiw:2003pm}. Similar to EdGB gravity, dCS gravity arises as an effective field theory from the compactification of  heterotic string theory~\cite{GREEN1984117,McNees:2015srl}. Such a theory is also important in the context of particle physics~\cite{Alexander:2009tp,Mariz:2004cv,MARIZ2008312,PhysRevD.78.025029}, loop quantum gravity~\cite{PhysRevD.80.104007,Taveras:2008yf}, and inflationary cosmology~\cite{Weinberg:2008hq}. Demanding that the critical length scale (below which higher curvature corrections beyond quadratic order cannot be neglected in the action) has to be smaller than the scale probed by table-top experiments, one finds $\sqrt{|\bar{\alpha}_{\DCS}|} < \mathcal{O}(10^8 \text{km})$~\cite{Yagi:2012ya}. Similar constraints have been placed from measurements of the frame-dragging effect by Gravity Probe B and LAGEOS satellites~\cite{AliHaimoud:2011fw}.

We now derive the expressions of the PPE parameters for dCS gravity. While BHs in EdGB gravity possess scalar monopole charges, BHs in dCS gravity possess scalar dipole charges which induce scalar quadrupolar emission~\cite{Yagi:2011xp}. On the other hand, scalar dipole charges induce a scalar interaction force between two BHs. Each BH also acquires a modification to the quadrupole moment away from the Kerr value. All of these modifications result in both dissipative and conservative corrections entering at the same order in gravitational waveforms. For spin-aligned binaries\footnote{See recent works~\cite{Loutrel:2018rxs,Loutrel:2018ydv} for precession equations in dCS gravity.}, corrections to Kepler's law and frequency evolution in dCS gravity are given in~\cite{Yagi:2012vf} within the slow-rotation approximation for BHs, from which we can derive
\begin{align}\label{dcs:gamma_r}
\gamma_r=&\frac{25}{256}\eta^{-9/5}\zeta_{\DCS}\chi_1 \chi_2 \nonumber \\ 
&-\frac{201}{3584}\eta^{-14/5}\zeta_{\DCS} \left(\frac{m_1^2}{m^2}\chi_2^2 +\frac{m_2^2}{m^2}\chi_1^2\right)
\end{align}
with $c_r=4$, and
  \begin{align}\label{dcs:gamma_fdot}
\gamma_{\dot{f}}=& \frac{11975}{12288}\eta^{-9/5}\zeta_{\DCS}\chi_1 \chi_2 \nonumber \\
&-\frac{96305}{172032 }\eta^{-14/5}\zeta_{\DCS} \left(\frac{m_1^2}{m^2}\chi_2^2+\frac{m_2^2}{m^2}\chi_1^2\right)\,.
 \end{align}
with $c_{\dot{f}}=4$. Here $\zeta_{\DCS}=16\pi \bar{\alpha}_{\text{dCS}}^2/m^4$ is the dimensionless coupling constant.
Using Eqs.~\eqref{dcs:gamma_r} and~\eqref{dcs:gamma_fdot} in Eqs.~\eqref{eq:amp-ppE-comparable} and~\eqref{eq:2v} respectively, one finds
  \begin{align}
 \label{eq:alpha}
 \alpha_{\DCS}=&\frac{57713}{344064}\eta ^{-14/5}\zeta_{\DCS} \left[-2 \delta_m \chi_a \chi_s \right. \nonumber\\ 
 &\left. +\left(1-\frac{14976 \eta }{57713}\right) \chi_a^2+\left(1-\frac{215876 \eta }{57713}\right) \chi_s^2\right]\,, \\
 \end{align}
with $a=4$, and
 \begin{align}
 \label{eq:beta}
 \beta_{\DCS}=&\frac{481525}{3670016}\eta ^{-14/5} \zeta_{\DCS} \left[-2 \delta_m \chi_a \chi_s \right.\nonumber\\ 
 &\left. +\left(1-\frac{4992 \eta }{19261}\right)\chi_a^2 +\left(1-\frac{72052 \eta }{19261}\right)\chi_s^2\right]\,.
 \end{align}
with $b=-1$. Here $\chi_{s,a}=(\chi_1 \pm \chi_2 )/2$ are the symmetric and antisymmetric combinations of dimensionless spin parameters 
and $\delta_m=(m_1-m_2)/m$ is the fractional difference in masses relative to the total mass. The above corrections enter at 2~PN order.

Can GW observations place stronger bounds on the theory? Current GW observations do not allow us to put any meaningful bounds on dCS gravity~\cite{Yunes:2016jcc} (see also~\cite{Yagi:2017zhb}). However, future observations have potential to place bounds on the theory that are six to seven orders of magnitude stronger than current bounds~\cite{Yagi:2012vf}. Such stronger bounds can be realized due to relatively strong gravitational field and large spins that source the pseudo-scalar field. Measuring GWs from extreme mass ratio inspirals with LISA can also place bounds that are three orders of magnitude stronger than current bounds~\cite{Canizares:2012is}.

 \subsection{Einstein-\AE ther  and Khronometric Theory}

In this section, we study two example theories that break Lorentz invariance in the gravity sector, namely Einstein-\AE ther and khronometric theory. Lorentz-violating theories of gravity are candidates for low-energy descriptions of quantum gravity~\cite{Blas:2014aca,Horava:2009uw}. Lorentz-violation in the gravity sector has not been as stringently constrained as that in the matter sector~\cite{Mattingly:2005re,Jacobson:2005bg,Liberati:2013xla} and several mechanisms exist that prevents percolation of the latter to the former~\cite{Liberati:2013xla,Pospelov:2010mp}.
 
Einstein-\AE ther theory is a vector-tensor theory of gravity, where along with the metric, a spacetime is endowed with a dynamical timelike unit vector (\AE ther) field~\cite{Jacobson:2000xp,Jacobson:2008aj}. Such a vector field specifies a particular rest frame at each point in spacetime, and hence breaks the local Lorentz symmetry. The amount of Lorentz violation is controlled by four coupling parameters 
$(c_1,c_2,c_3,c_4)$. Einstein-\AE ther theory preserves diffeomorphism invariance and hence is a Lorentz-violating theory without abandoning the framework of GR~\cite{Jacobson:2008aj}. 
Along with the spin-2 gravitational perturbation of GR, the theory predicts the existence of the spin-1 and spin-0 perturbations~\cite{Foster:2006az,Jacobson:2004ts,PhysRevD.76.084033}. Such perturbation modes propagate at speeds that are functions of the coupling parameters $c_i$, and in general differ from the speed of light~\cite{Jacobson:2004ts}.

Khronometric theory is a variant of Einstein-\AE ther theory, where the $\AE$ther field is restricted to be hypersurface-orthogonal. Such a theory arises as a low-energy limit of Ho\v{r}ava gravity, a power-counting renormalizable quantum gravity model with only spatial diffeomorphism invariance~\cite{Blas:2009qj,Berti:2015itd,Horava:2009uw,Nishioka:2009iq,Visser:2009fg}. The amount of Lorentz violation in the theory is controlled by three parameters, $(\bar \alpha_{\KG}, \bar \beta_{\KG}, \bar \lambda_{\KG})$. Unlike Einstein-\AE ther theory, the spin-1 propagating modes are absent in khronometric theory.

Most of parameter space in Einstein-\AE ther and khronometric theory have been constrained stringently from current observations and theoretical requirements. Using the measurement of the arrival time difference between GWs and electromagnetic waves in GW170817, the difference in the propagation speed of GWs away from the speed of light has been constrained to be less than $\sim 10^{-15}$~\cite{TheLIGOScientific:2017qsa,Monitor:2017mdv}. Such a bound can be mapped to bounds on Lorentz-violating gravity as $|c_1 + c_3| \lesssim 10^{-15}$~\cite{Gong:2018cgj,Oost:2018tcv} and $|\bar{\beta}_{\KG} | \lesssim10^{-15}$~\cite{Gumrukcuoglu:2017ijh}\footnote{Such bounds are consistent with the prediction in~\cite{Hansen:2014ewa} based on~\cite{Nishizawa:2014zna}.}. Imposing further constraints from solar system experiments~\cite{Bailey:2006fd,Foster:2005dk,Will:2005va}, Big Bang nucleosynthesis~\cite{Audren:2013dwa} and theoretical constraints such as the stability of propagating modes, positivity of their energy density~\cite{Eling:2005zq} and the absence of gravitational Cherenkov radiation\cite{Elliott:2005va},  allowed regions in the remaining parameter space have been derived for Einstein-\AE ther~\cite{Oost:2018tcv} and khronometric~\cite{Gumrukcuoglu:2017ijh} theory. Binary pulsar bounds on these theories were studied in~\cite{Yagi:2013ava,Yagi:2013qpa} before the discovery of GW170817, within a parameter space that is different from the allowed regions in~\cite{Oost:2018tcv,Gumrukcuoglu:2017ijh}.

Let us now derive the PPE parameters in Einstein-\AE ther and khronometric theories.  Propagation of the scalar and vector modes is responsible for dipole radiation and loss of angular momentum in binary systems, which increase the amount of orbital decay rate. 
Regarding Einstein-\AE ther theory, the PPE phase correction is given by~\cite{Hansen:2014ewa} 
 \begin{align}
 \label{eq:phase-EA}
 \beta_{\EA}=&-\frac{5}{3584}\eta ^{2/5} \frac{(s_1^\EA-s_2^\EA)^2}{[(1-s_1^\EA) (1-s_2^\EA)]^{4/3}}\nonumber\\
 & \times \frac{(c_{14}-2) w_0^3-w_1^3}{c_{14} w_0^3 w_1^3}
 \end{align}
with $b=-7$. Here $w_s$ is the propagation speed of the spin-$s$ modes in Einstein-\AE ther theory given by~\cite{Jacobson:2008aj}  
\ba \label{eq:Prop_Speed_EA_1}
w_0^2 &=& \frac{(2-c_{14}) c_{123}}{(2+3c_2+c_{+}) (1-c_{+}) c_{14}}\,, \\ \label{eq:Prop_Speed_EA_2}
w_1^2 &=& \frac{2 c_1 - c_{+} c_{-}}{2(1-c_{+}) c_{14}}\,, \\  \label{eq:Prop_Speed_EA_3}
w_2^2 &=& \frac{1}{1 - c_+}\,,
\ea
with
\be
c_{14}\equiv c_1+c_4\,, \quad c_{\pm} \equiv c_1 \pm c_3\,, \quad c_{123} \equiv c_1 + c_2 + c_3\,. 
\ee
$s_A$ in Eq.~\eqref{eq:phase-EA} is the sensitivity of the $A$-th body and has been calculated only for NSs~\cite{Yagi:2013ava,Yagi:2013qpa}.
Given that the leading order correction in Einstein-\AE ther theory arises from the dissipative sector~\cite{Hansen:2014ewa}, we can use Eq.~\eqref{eq:2w2} to find the PPE amplitude correction as\footnote{Eqs.~\eqref{eq:amp-EA} and~\eqref{eq:amp-KG} correct errors in~\cite{Hansen:2014ewa}.}
 \begin{align}
 \label{eq:amp-EA}
 \alpha_{\EA}=&-\frac{5}{96}\eta ^{2/5} \frac{(s_1^\EA-s_2^\EA)^2}{[ (1-s_1^\EA) (1-s_2^\EA)]^{4/3}} \nonumber\\
 & \times  \frac{(c_{14}-2) w_0^3-w_1^3}{c_{14} w_0^3 w_1^3}
 \end{align}
 with $a=-2$.  Similar to Einstein-\AE ther theory, the PPE parameters in khronometric theory is given by~\cite{Hansen:2014ewa}
 \begin{align}
 \beta_{\KG}=&-\frac{5}{3584}\eta ^{2/5}\frac{(s_1^\KG-s_2^\KG)^2}{[(1-s_1^\KG) (1-s_2^\KG)]^{4/3}} \nonumber\\
 & \times \sqrt{\bar \alpha_{\KG}}\left[\frac{(\bar \beta_{\KG}-1)(2+\bar \beta_{\KG}+3\bar \lambda_{\KG})}{(\bar \alpha_{\KG}-2)(\bar \beta_{\KG}+\bar \lambda_{\KG})}\right]^{3/2}
 \end{align}
 with $b=-7$, and
 \begin{align}
 \label{eq:amp-KG}
 \alpha_{\KG}=&-\frac{5}{96} \eta ^{2/5}\frac{(s_1^\KG-s_2^\KG)^2}{[(1-s_1^\KG) (1-s_2^\KG)]^{4/3}} \nonumber\\
 & \times\sqrt{\bar \alpha_{\KG}} \left[\frac{(\bar \beta_{\KG}-1)(2+\bar \beta_{\KG}+3\bar \lambda_{\KG})}{(\bar \alpha_{\KG}-2)(\bar \beta_{\KG}+\bar \lambda_{\KG})}\right]^{3/2}
 \end{align}
 with $a=-2$. These corrections enter at $-1$~PN order.
 
Above corrections to the gravitational waveform can be used to compute current and projected future bounds on the theories with GW observations, provided one knows what the sensitivities are for compact objects in binaries. Unfortunately, such sensitivities have not been calculated for BHs, and hence, one cannot derive bounds on the theories from recent binary BH merger events. Instead, Ref.~\cite{Yunes:2016jcc} used the next-to-leading 0~PN correction that is independent of the sensitivities and derived bounds from GW150914 and GW151226, though such bounds are weaker than those from binary pulsar observations~\cite{Yagi:2013ava,Yagi:2013qpa}. On the other hand, Ref.~\cite{Hansen:2014ewa} includes both the leading and next-to-leading corrections to the waveform and estimate projected future bounds with GWs from binary NSs. The authors found that bounds from second-generation ground-based detectors are less stringent than existing bounds even with their design sensitivities.
However, third-generation ground-based ones and space-borne interferometers can place constraints that are comparable, and in some cases, two orders of magnitude stronger compared to the current bounds~\cite{Chamberlain:2017fjl,Hansen:2014ewa}.

 \subsection{Noncommutative Gravity}
Although the concept of nontrivial commutation relations of spacetime coordinates is rather old~\cite{Snyder:1946qz,Snyder:1947nq}, the idea has revived recently with the development of noncommutative geometry~\cite{connes1985non,connes1995noncommutative,1987117,Landi:1997sh,Woronowicz:1987vs}, and the emergence of noncommutative structure of spacetime in a specific limit of string theory~\cite{WITTEN1986253,Seiberg:1999vs}. Quantum field theories on noncommutative spacetime have been studied extensively as well~\cite{Douglas:2001ba,Rivelles:2002ez,Szabo:2001kg}. In the simplest model of  noncommutative gravity, spacetime coordinates are promoted to operators, which satisfy a canonical commutation relation:
\be
\left[\hat{x}^{\mu},\hat{x}^{\nu}\right]=i \theta^{\mu\nu}\,,
\ee
where $\theta^{\mu\nu}$ is a real constant antisymmetric tensor. In ordinary quantum mechanics, Planck's constant $\hbar$ measures the quantum fuzziness of phase space coordinates. In a similar manner, $\theta^{\mu\nu}$ introduces a new fundamental scale which measures the quantum fuzziness of spacetime coordinates~\cite{Kobakhidze:2016cqh}.

In order to obtain stringent constraints on the scale of noncommutativity, low-energy experiments are advantageous over high-energy ones~\cite{Carroll:2001ws,Mocioiu:2000ip}. Low-energy precision measurements such as clock-comparison experiments with nuclear-spin-polarized $_{}^{9}\textrm{Be}^+$ ions~\cite{PhysRevLett.54.2387} give a constraint on noncommutative scale as $1/\sqrt{\theta}\gtrsim 10$ TeV~\cite{Carroll:2001ws}, where $\theta$ refers to the magnitude of the spatial-spatial components of $\theta^{\mu\nu}$\footnote{The corresponding bound on the time-spatial components of $\theta^{\mu\nu}$ is roughly six orders magnitude weaker than that on the spatial-spatial components.}. A similar bound has been obtained from the measurement of the Lamb shift~\cite{PhysRevLett.86.2716}. Another speculative bound is derived from the analysis of atomic experiments which is 10 orders of magnitude stronger~\cite{Berglund:1995zz,Mocioiu:2000ip}. Study of inflationary observables using cosmic microwave background data from Planck gives the lower bound on the energy scale of noncommutativity as  19 TeV~\cite{calmet2015inflation,PhysRevD.91.083503}.

Let us now review how the binary evolution is modified from that in GR in this theory.
Several formulations of noncommutative gravity exist~\cite{Aschieri:2005yw,Aschieri:2005zs,Calmet:2005qm,Chamseddine:2000si,Kobakhidze:2006kb,Szabo:2006wx}, though the first order noncommutative correction vanishes in all of them~\cite{Calmet:2006iz,Mukherjee:2006nd} and the leading order correction enters at second order. On the other hand, first order  corrections may arise from gravity-matter interactions~\cite{Kobakhidze:2007jn,Mukherjee:2006nd}. Thus one can neglect corrections to the pure gravity sector and focus on corrections to the matter sector (i.e., energy-momentum tensor)~\cite{Kobakhidze:2016cqh}. Making corrections to classical matter source and following an effective field theory approach, expressions of energy and GW luminosity for quasi-circular BH binaries have been derived in Ref.~\cite{Kobakhidze:2016cqh}, which give the correction to the frequency evolution in Eq.~\eqref{eq:2m} as
\ba \label{noncom:gamma_fdot}
\gamma_{\dot{f}}=\frac{5}{4}\eta ^{-4/5}(2 \eta -1)\Lambda^2
\ea
with $c_{\dot{f}}=4$ and $\Lambda^2 = \theta^{0i} \theta_{0i}/(l_p^2 t_p^2)$ with $l_p$ and $t_p$ representing the Planck length and time respectively. On the other hand, modified Kepler's law in Eq.~\eqref{eq:2k} can be found as~\cite{Kobakhidze:2016cqh}
\be \label{noncom:gamma_r}
\gamma_r=\frac{1}{8}\eta ^{-4/5}(2 \eta -1)\Lambda^2
\ee
with $c_r=4$. 

We are now ready to derive the PPE parameters in noncommutative gravity.
Given that the dissipative and conservative leading corrections enter at the same PN order, one can use Eqs.~\eqref{noncom:gamma_fdot} and~\eqref{noncom:gamma_r} in Eq.~\eqref{eq:amp-ppE-comparable} to find the PPE amplitude correction as
\begin{equation}
 \alpha_{\NC}=-\frac{3}{8}\eta ^{-4/5}(2 \eta -1) \Lambda ^2
 \end{equation}
with $a=4$. Similarly, substituting Eq.~\eqref{noncom:gamma_fdot} into Eq.~\eqref{eq:2v} gives the PPE phase correction as
 \begin{equation}
 \beta_{\NC}=-\frac{75}{256}\eta ^{-4/5}(2 \eta -1) \Lambda ^2
 \end{equation}
with $b=-1$. $\beta_{\NC}$ can also be read off from the phase correction derived in~\cite{Kobakhidze:2016cqh}.
The above corrections enter at 2~PN order. 

The above phase correction has already been used to derive bounds on noncommutative gravity from GW150914 as $\sqrt{\Lambda} \lesssim 3.5$~\cite{Kobakhidze:2016cqh}, which means that the energy scale of noncommutativity has been constrained to be the order of the Planck scale. Such a bound, so far, is the most stringent constraint on noncommutative scale and is 15 orders of magnitude stronger compared to the bounds coming from particle physics and low-energy precision measurements\footnote{Notice that the GW bound is on the time-spatial components of $\theta^{\mu\nu}$, while most of particle physics and low-energy precision experiments place bounds on its spatial-spatial components.}.
 
 \section{Varying-$G$ Theories}\label{gdot}
 
Many of the modified theories of gravity that violate the strong equivalence principle~\cite{DiCasola:2013iia,Will:2014kxa,0264-9381-7-10-007} predict that locally measured gravitational constant ($G$) may vary with time~\cite{uzan:2010pm}. Since the gravitational self-energy of a body is a function of the gravitational constant, in a theory where $G$ is time-dependent, masses of compact bodies are also time-dependent~\cite{PhysRevLett.65.953}. The rate at which the mass of an object varies with time is proportional to the rate of change of the gravitational coupling constant~\cite{PhysRevLett.65.953}. Such a variation of mass, together with the conservation of linear momentum, causes  compact bodies to experience anomalous acceleration, which results in a time-evolution of the specific angular momentum~\cite{PhysRevLett.65.953}. Existing experiments that search for variations in $G$ at present time (i.e., at very small redshift) include lunar laser ranging observations~\cite{Williams:2004qba}, pulsar timing observations~\cite{Deller:2008jx,Kaspi:1994hp}, radar observations of planets and spacecraft~\cite{Pitjeva2005}, and surface temperature observations of PSR J0437-4715~\cite{Jofre:2006ug}. Another class of constraints on a long-term variation of $G$ comes from Big Bang nucleosynthesis~\cite{Bambi:2005fi,Copi:2003xd} and helioseismology~\cite{0004-637X-498-2-871}. The most stringent bound on $|\dot{G}/G|$ is of the order $\lesssim 10^{-14} \,\mathrm{yr}^{-1}$~\cite{2018NatCo...9..289G}.

More than one gravitational constants can appear in different areas of a gravitational theory. Here we introduce two different kinds of gravitational constant, one that arises in the dissipative sector and another in the conservative sector. The constant which enters in the GW luminosity through Einstein equations, i.e. the constant in Eq.~\eqref{eq:2h2}, is the one we refer to as dissipative gravitational constant ($G_D$), while that enters in Kepler's law or binding energy of the binary is what we refer as the conservative one ($G_C$). These two constants are the same in GR, but they can be different in some modified theories of gravity. An example of such a theory is Brans-Dicke theory with a cosmologically evolving scalar field~\cite{Will2006}. 

The PPE parameters for varying-$G$ theories have previously been derived in~\cite{Yunes:2009bv} for $G_D = G_C$. Here, we improve the analysis by considering the two different types of gravitational constant and including variations in masses, which are inevitable for strongly self-gravitating objects when $G$ varies~\cite{PhysRevLett.65.953}.
We also correct small errors in~\cite{Yunes:2009bv}. We follow the analysis of~\cite{Yagi:2011yu} that derives gravitational waveforms from BH binary inspirals with varying mass effects from the specific angular momentum. We also present another derivation in App.~\ref{appendix_2} using the energy balance argument in~\cite{Yunes:2009bv}.

The formalism presented in Sec.~\ref{section:ppE} assumes that $G$ and the masses to be constant, and hence are not applicable to varying-$G$ theories. Thus, we will derive the PPE parameters in varying-$G$ theories by promoting the PPE formalism to admit time variation in the gravitational constants and masses as
 \begin{eqnarray}\label{eq:3.7a2}
 m_A(t)\approx m_{A,0}+\dot{m}_{A,0}(t-t_0)\,, \\
   \label{eq:3.7a4}  G_C(t)\approx  G_{C,0}+\dot{G}_{C,0}(t-t_0)\,, \\
   \label{eq:3.7a4-2}  G_D(t)\approx  G_{D,0}+(1+\delta_\Gdot)\dot{G}_{C,0}(t-t_0)\, , 
 \end{eqnarray}
where $t_0$ is the time of coalescence. Here we assumed that spatial variations of $G_C$ and $G_D$ are small compared to variations in time. $\delta_\Gdot$ gives the fractional difference between the rates at which $G_C$ and $G_D$ vary with time, and could be a function of parameters in a theory. The subscript $0$ denotes that the quantity is measured at the time $t=t_0$. 
Other time variations to consider are those in the specific angular momentum $j$ and the total mass $m$:
\ba
\label{eq:3.7a3-2}  j(t)\approx j_0+\dot{j}_0(t-t_0)\,, \\
m(t)\approx m_0+\dot{m}_0(t-t_0)\,.
 \ea
$\dot{j}_0$ and $\dot{m_0}$ can be written in terms of binary masses and sensitivities defined by
\begin{equation}
s_A=-\frac{G_C}{m_A}\frac{\delta m_A}{\delta  G_C}\,,
\end{equation}
as~\cite{PhysRevLett.65.953}
\begin{eqnarray}\label{eq:3.7a4-3}
\dot{j}_0&=&\frac{m_{1,0}s_{1,0}+m_{2,0}s_{2,0}}{m_{1,0}+m_{2,0}}\frac{\dot{G}_{C,0}}{G_{C,0}}j_0\,, \\
\label{eq:3.7a4-4}
\dot{m}_0&=&-\frac{m_{1,0}s_{1,0}+m_{2,0}s_{2,0}}{m_{1,0}+m_{2,0}}\frac{\dot{G}_{C,0}}{G_{C,0}}m_0\,,
\end{eqnarray}
respectively. 

Next, we explain how the binary evolution is affected by the variation of the above parameters. 
First, GW emission makes the orbital separation $r$ decay with the rate given by~\cite{PhysRevD.49.2658}
 \begin{equation}
 \dot{r}_{\GW}=-\frac{64}{5}\frac{G_D G_C^2 \mu m^2}{r^3}\,.
 \end{equation}
Second, time variation of the total mass, (conservative) gravitational constant and specific angular momentum changes $r$ at a rate of
 \begin{equation}
 \dot{r}_\Gdot=-\left(\frac{\dot{G}_{C,0}}{G_C}+\frac{\dot{m}_0}{m}-2\frac{\dot{j}_0}{j}\right)r\,,
 \end{equation}
 which is derived by taking a time derivative of the specific angular momentum $j\equiv\sqrt{G_Cmr}$. Having the evolution of $r$ at hand, one can derive the evolution of the orbital angular frequency using Kepler's third law as
 \begin{equation}\label{eq:3.7a}
 \dot{\Omega}=\frac{1}{2\Omega r^3}\left(m\dot{G}_{C,0}+\dot{m}_0G_C-3mG_C\frac{\dot{r}}{r}\right)\,.
 \end{equation}
Using the evolution of the binary separation $\dot{r}\equiv \dot{r}_{\GW}+\dot{r}_\Gdot$ in Eq.~\eqref{eq:3.7a}, together with Eqs.~\eqref{eq:3.7a4-3} and~\eqref{eq:3.7a4-4},  we can find the GW frequency evolution as
\allowdisplaybreaks
\begin{eqnarray} 
\label{eq:3.7b}
 \dot{f}&=&\frac{\dot{\Omega}}{\pi} \nonumber \\
 &=&\frac{96}{5}\pi^{8/3}G_C^{2/3}G_D\mathcal{M}^{5/3}f^{11/3}\left\{1 \right. \nonumber \\
 && \left. +\frac{5}{96}\frac{\dot{G}_{C,0} G_C}{G_D\eta}[2m-5(m_{1,0}s_{1,0}+m_{2,0}s_{2,0})] x^{-4}  \right\}\,, \nonumber \\
\end{eqnarray}
where $x\equiv (\pi G_C  m f)^{2/3}$ is the squared velocity of the relative motion. Here we only considered the leading correction to the frequency evolution entering at $-4$~PN order. Using Eqs.~\eqref{eq:3.7a2}--\eqref{eq:3.7a4-2} and~\eqref{eq:3.7a4-4} into Eq.~\eqref{eq:3.7b}, one finds
\begin{align}\label{eq:3.7b2}
\dot{f}&=\frac{96}{5} \pi ^{8/3}\, f^{11/3}\,\eta_0 \,G_{C,0}^{2/3}\,G_{D,0}\,m_0^{5/3}\left\{1 \right.\nonumber\\& \left.-\frac{5\,G_{C,0}\,\dot{G}_{C,0}}{768\, \eta_0\, G_{D,0}^2}\left[3(1+\delta_\Gdot)G_{C,0}m_0-(3s_{1,0}+3s_{2,0}\right.\right.\nonumber\\ &\left.\left.+14)G_{D,0}m_0+41(m_{1,0}s_{1,0}+m_{2,0}s_{2,0})G_{D,0}\right]x_0^{-4}\right\}\,.
\end{align}
Notice that $G_{C,0}$ and $G_{D,0}$ differ only by a constant quantity, and such a difference will enter in $\dot f$ at 0~PN order which is much higher than the $-4$~PN corrections. We will thus ignore such 0~PN corrections and simply use $G_{D,0}= G_{C,0}\equiv G_0$ from now on. 

Based on the above binary evolution, we now derive corrections to the GW phase.
We integrate Eq.~\eqref{eq:3.7b2} to obtain time before coalescence $t(f)-t_0$ and the GW phase $\phi(f)\equiv \int 2 \pi f dt = \int (2\pi f/\dot{f})df$ as
\allowdisplaybreaks
\begin{eqnarray}
t(f)&=&t_0 -\frac{5}{256}G_0 \mathcal{M}_0{u_0}^{-8}\left\{1 \right. \nonumber \\ 
&& \left. -\frac{5}{1536}\frac{\dot{G}_{C,0}}{\eta_0}\left[11m_0+3(s_{1,0}+s_{2,0}-\delta_\Gdot)m_0 \right. \right. \nonumber \\
&& \left. \left. -41 (m_{1,0}s_{1,0}+m_{2,0}s_{2,0})\right]x_0^{-4} \right\}\,, \\
\phi(f)&=&\phi_0  -\frac{1}{16}{u_0}^{-5}\left\{1  \right. \nonumber \\ 
&& \left.  -\frac{25}{9984}\frac{\dot{G}_{C,0}}{\eta_0}\left[11m_0+3(s_{1,0}+s_{2,0}-\delta_\Gdot)m_0 \right. \right. \nonumber \\
&& \left. \left. -41 (m_{1,0}s_{1,0}+m_{2,0}s_{2,0})\right]x_0^{-4} \right\}\,,
\end{eqnarray}
with $u_0 \equiv (\pi G_0 \mathcal{M}_0 f)^\frac{1}{3}$.
The GW phase in the Fourier space is then given by
\bw
\begin{align}\label{eq:3.7c}
\Psi(f)=&2\pi ft(f)-\phi(f)-\frac{\pi}{4}\nonumber\\
=&2\pi f t_0-\phi_0-\frac{\pi}{4}+\frac{3}{128}u_0^{-5}\left\{1-\frac{25}{19968}\frac{\dot{G}_{C,0}}{\eta_0}\left[11m_0+3(s_{1,0}+s_{2,0}-\delta_\Gdot)m_0-41 (m_{1,0}s_{1,0}+m_{2,0}s_{2,0})\right]x_0^{-4} \right\}\,.
\end{align}
\ew
From Eq.~\eqref{eq:3.7c}, one finds the PPE phase parameters as $b=-13$ and
 \begin{eqnarray}\label{eq:beta_gdot}
 \beta_{\dot{G}}&=&-\frac{25}{851968}\dot{G}_{C,0}\,\eta_0^{3/5}\left[11m_0 +3(s_{1,0}+s_{2,0}-\delta_\Gdot)m_0 \right. \nonumber \\
 && \left. -41 (m_{1,0}s_{1,0}+m_{2,0}s_{2,0})\right]\,. 
  \end{eqnarray}

Next, we derive the PPE amplitude parameters.
Using Kepler's law to Eq.~\eqref{eq:2e}, one finds
\begin{eqnarray}\label{eq:3.7d}
\tilde{\mathcal{A}}(f)&\propto&\frac{1}{\sqrt{\dot{f}}}\frac{G_D(t)}{D_L}\mu(t) r(t)^2f^2 \nonumber \\
&\propto&\frac{1}{\sqrt{\dot{f}}}G_D(t){G_C(t)}^{2/3}\mu(t){m(t)}^{2/3} \,.
\end{eqnarray} 
Using further Eqs.~\eqref{eq:3.7a2}--\eqref{eq:3.7a4-2} in Eq.~\eqref{eq:3.7d}, we find the amplitude PPE parameters as $a = -8$ and
\begin{eqnarray}
\label{eq:3.7d2}
 \alpha_{\dot{G}}&=&\frac{5}{512}\eta_0^{3/5}\dot{G}_{C,0}\left[-7m_0+(s_{1,0}+s_{2,0}-\delta_\Gdot)m_0 \right. \nonumber \\
 && \left.+13 (m_{1,0}s_{1,0}+m_{2,0}s_{2,0})\right]\,.
 \end{eqnarray}

Let us comment on how the above new PPE parameters in varying-$G$ theories differ from those obtained previously in~\cite{Yunes:2009bv}. The latter considers $G_D = G_C$ (which corresponds to $\delta_\Gdot = 0$) and $s_A = 0$ (which is only true for weakly-gravitating objects). However, the above expressions for the PPE parameters do not reduce to those in~\cite{Yunes:2009bv} under these limits. This is because Ref.~\cite{Yunes:2009bv} did not take into account the fact that the binding energy is not conserved in the absence of GW emission in varying-$G$ theories. In App.~\ref{appendix_2}, we show that the correct application of the energy balance law does indeed lead to the same conclusion as in this section.

Eqs.~\eqref{eq:beta_gdot} and~\eqref{eq:3.7d2} can be used to constrain varying-$G$ theories with GW observations. Recent GW events (GW150914 and GW151226) place constraints on variation of G which are much weaker than the current constraints~\cite{Yunes:2016jcc}. 
Projected GW bounds have been calculated in Ref.~\cite{Yunes:2009bv} (see~\cite{Chamberlain:2017fjl} for an updated forecast of future GW bounds on $\dot G$) which gives $|\dot{G}_0/G_0| \lesssim 10^{-11}\, \mathrm{yr}^{-1}$, considering a single merger event. Although GW bounds are less stringent compared to the existing bounds~\cite{Will2006}, they are unique in the sense that they can provide constraints at intermediate redshifts, while the existing bounds are for very small and large redshifts~\cite{Yunes:2009bv}. Furthermore, GW constraints give $\dot{G}_0/G_0$ at the location of merger events, which means that a sufficient number of GW observations can be used to construct a 3D constraint map of $\dot{G}_0/G_0$ as a function of sky locations and redshifts~\cite{Yunes:2009bv}.

\section{Conclusions}\label{conclusions}

We derived non-GR corrections to the GW phase and amplitude in various modified theories of gravity. We achieved this by revisiting the standard PPE formalism and considered generic corrections to the GW frequency evolution and Kepler's third law that have been derived in many non-GR theories. Such a formalism yields the expressions of the PPE parameters which are simpler compared to the original formalism~\cite{Yunes:2009ke,Chatziioannou:2012rf}. 
We derived the PPE amplitude parameters for the first time in EdGB, dCS and noncommutative gravity. 
We also corrected some errors in the expressions of the PPE amplitude parameters in Einstein-\AE ther and khronometric theories in previous literature~\cite{Hansen:2014ewa}.

We also considered the PPE formalism with variable gravitational constants by extending previous work~\cite{Yunes:2009bv} in a few different ways. One difference is that we introduced two different gravitational constants, one entering in the GW luminosity (dissipative $G$) and the other entering in the binding energy and Kepler's law (conservative $G$). We also included time variations of component masses in a binary in terms of the sensitivities following~\cite{PhysRevLett.65.953}, which is a natural consequence in varying-$G$ theories. We further introduced the effect of non-conservation of the binding energy in the energy balance law. Such an effect arises due to an anomalous acceleration caused by time variations in $G$ or masses~\cite{PhysRevLett.65.953} that was not accounted for in the original work of~\cite{Yunes:2009bv}.  Including all of these, we derived the PPE amplitude and phase corrections to the gravitational waveform from compact binary inspirals.

The analytic expressions of the PPE corrections derived in this paper, especially those in the amplitude, can be used to improve analyses on testing GR with observed GW events and to derive new projected bounds with future observations, since most of previous literature only include phase corrections. For example, one can reanalyze the available GW data for testing GR including amplitude corrections with a Bayesian analysis~\cite{TheLIGOScientific:2016src}. One can also carry out a similar Fisher analysis as in~\cite{Yunes:2016jcc} by including amplitude corrections and mapping bounds on generic parameters to those on fundamental pillars in GR. GW amplitude corrections are also crucial for testing strong-field gravity with astrophysical stochastic GW backgrounds~\cite{Maselli:2016ekw,Yagi:2017zhb}. One could further improve the analysis presented in this paper by considering binaries with eccentric orbits~\cite{Loutrel:2014vja} or including spin precession~\cite{Huwyler:2014gaa,Loutrel:2018ydv}.
We leave these possible avenues of extensions for future work.

\label{sec:conclusions}

\acknowledgments
We would like to thank Nicol\' as Yunes for fruitful discussions.
K.Y. would like to acknowledge networking support by the COST Action GWverse CA16104.

 \appendix 
 
 \section{Original PPE Formalism}
 \label{appendix}

In this appendix, we review the original PPE formalism. In particular, we will show how the amplitude and phase corrections depend on  conservative and dissipative corrections, where the former are corrections to the effective potential of a binary while the latter are those to the GW luminosity. We will mostly follow the analysis in~\cite{Chatziioannou:2012rf}.

First, let us introduce conservative corrections. We modify the reduced effective potential of a binary as
 \begin{equation}\label{eq:h}
 V_{\text{eff}}=\left(-\frac{m}{r}+\frac{{L}^2_{z}}{2\mu^2r^2}\right)\left[1+A \left(\frac{m}{r}\right)^p\right]\,,
 \end{equation}
where $L_{z}$ is the $z$-component of the angular momentum. $A$ and $p$ show the magnitude and exponent of the non-GR correction term respectively. Such a modification to the effective potential also modifies Kepler's law. Taking the radial derivative of $V_{\text{eff}}$ in Eq.~\eqref{eq:h} and equating it to zero gives modified Kepler's law as
 \begin{equation}
 \Omega^2=\frac{m}{r^3} \left[1+\frac{1}{2} A \, p\left(\frac{m}{r}\right)^p\right]\,.
 \end{equation}
 The above equation further gives the orbital separation as
 \begin{equation}\label{eq:g}
 r=r_{\GR}\left[1+\frac{1}{6}A\, p\, \eta^{-\frac{2p}{5}}u^{2p}\right]\,,
 \end{equation}
where to leading PN order, $r_{\GR}$ is given by Kepler's law as $r_{\GR}=(m/\Omega^2)^{1/3}$. For a circular orbit, radial kinetic energy does not exist and the effective potential energy is same as the binding energy of the binary. Using Eq.~\eqref{eq:g} in Eq.~\eqref{eq:h} and keeping only to leading order in non-GR corrections, the binding energy becomes
\begin{align}\label{eq:j}
E=-\frac{1}{2}\eta^{-2/5}u^2\left[1-\frac{1}{3}A(2p-5)\eta^{-\frac{2p}{5}}u^{2p}\right]\,.
\end{align}

Next, let us introduce dissipative corrections. Such corrections to the GW luminosity can be parameterized by
 \begin{equation}\label{eq:a}
 \dot{E}=\dot{E}_{\GR}\left[1+B\left(\frac{m}{r}\right)^q\right]\,,
 \end{equation}
 where $\dot{E}_{GR}$ is the GR luminosity which is proportional to $v^2(m/r)^4$ with $v=r\,\Omega  = (\pi m f)^{1/3}$ representing the relative velocity of binary components\footnote{If we assume $\dot{E}_{\GR}$ to be proportional to $r^4\, \Omega^6$ which directly follows from the quadrupole formula without using Kepler's law, we will find slightly different expressions for $\dot{f}$ and the waveform~\cite{Chatziioannou:2012rf}}. 
 
 Let us now derive the amplitude corrections. First, using Eqs.~\eqref{eq:j} and~\eqref{eq:a} and applying the chain rule, the GW frequency evolution is given by
 \begin{align}\label{eq:f}
 \dot{f}&=\frac{df}{dE}\frac{dE}{dt}\nonumber\\ &=\dot{f}_{GR}\left[1+B\eta^{-\frac{2q}{5}} u^{2q}+\frac{1}{3}A(2p^2-2p-3)\eta^{-\frac{2p}{5}}u^{2p} \right]\,,
 \end{align}
where $\dot{f}_{\GR}$ is given by Eq.~\eqref{eq:2s}. 
Next, using Eqs.~\eqref{eq:g} and~\eqref{eq:f} to Eq.~\eqref{eq:2e} and keeping only to leading order in non-GR corrections, the GW amplitude in Fourier domain becomes
\begin{align}\label{eq:o2}
\tilde{\mathcal{A}}(f)=\tilde{\mathcal{A}}_{GR} \left[1-\frac{B}{2}\eta^{-\frac{2q}{5}}u^{2q}-\frac{1}{6}A(2p^2-4p-3)\eta^{-\frac{2p}{5}}u^{2p}\right]\,.
\end{align}

Next, we move onto deriving phase corrections. One can derive the GW phase in Fourier domain by integrating Eq.~\eqref{eq:2r} twice. Equivalently, one can use the following expression
\begin{equation}\label{eq:n}
\Psi(f)=2\pi f t(f)-\phi(f)-\frac{\pi}{4}\,,
\end{equation}
where $t(f)$ gives the relation between time and frequency and can be obtained by integrating~\eqref{eq:f} as
\begin{align}\label{eq:l}
t(f)=&\int \frac{dt}{df} \, df\nonumber\\=&t_0 -\frac{5 \mathcal{M}}{256 u^8}\left[1+\frac{4}{3}A\frac{\left(2 p^2-2 p-3\right)}{(p-4)}\eta ^{-\frac{2 p}{5}} u^{2 p}\right.\nonumber\\ &\left. +\frac{4}{q-4}B\eta ^{-\frac{2q}{5}} u^{2q}\right]\,,
\end{align}
with $t_0$ representing the time of coalescence and keeping only the Newtonian term and leading order non-GR corrections. On the other hand, $\phi(f)$ in Eq.~\eqref{eq:n} corresponds to the GW phase in time domain and can be calculated from Eq.~\eqref{eq:f} as
\begin{align}\label{eq:m}
\phi(f)=&\int 2 \pi f dt=\int\frac{2\pi f}{\dot{f}}df\nonumber\\
=&\phi_0 -\frac{1}{16 u^5}\left[1+\frac{5}{3}A\frac{\left(2 p^2-2 p-3\right)}{(2 p-5)} \eta ^{-\frac{2 p}{5}} u^{2 p}\right. \nonumber \\
& \left. +\frac{5}{2 q-5}B\eta ^{-\frac{2 q}{5}} u^{2 q}\right]\,,
\end{align}
with $\phi_0$ representing the coalescence phase. Using Eqs.~\eqref{eq:l} and~\eqref{eq:m} into~\eqref{eq:n} and writing $\Psi(f)$ as $\Psi_{\GR}(f)+\delta\Psi(f)$, non-GR modifications to the phase can be found as
 \begin{align}\label{eq:p2}
\delta\Psi(f)=&-\frac{5}{32}A\frac{2p^2-2p-3}{(4-p)(5-2p)}\eta^{-\frac{2p}{5}}u^{2p-5}\nonumber\\ &-\frac{15}{32}B\frac{1}{(4-q)(5-2q)}\eta^{-\frac{2q}{5}}u^{2q-5}\,,
 \end{align}
with $\Psi_{\GR}$ to leading PN order is given by~\cite{Blanchet:1995ez}
\begin{equation}
\label{eq:Psi_GR}
\Psi_{\GR}=2\pi f t_0-\phi_0-\frac{\pi}{4}+\frac{3}{128}u^{-5}\,.
\end{equation}

We can easily rewrite the above expressions using $\gamma_r$ and $c_r$. 
Comparing Eq.~\eqref{eq:g} with Eq.~\eqref{eq:2k}, we find
\begin{equation}\label{eq:i}
A=\frac{12\gamma_r}{c_r}\eta^{\frac{c_r}{5}}\,, \quad p=\frac{c_r}{2}\,.
\end{equation}
Using this, we can rewrite the GW amplitude in Eq.~\eqref{eq:o2} as
\begin{align}\label{eq:o}
\tilde{\mathcal{A}}(f)=\tilde{\mathcal{A}}_{GR} \left[1-\frac{B}{2}\eta^{-\frac{2q}{5}}u^{2q}-\frac{\gamma_r}{c_r}(c^2_r-4c_r-6)u^{c_r}\right]\,.
\end{align}
Similarly, one can rewrite the correction to the GW phase in Eq.~\eqref{eq:p2} as
\begin{align}\label{eq:p}
\delta\Psi(f)=&-\frac{15}{8}\frac{\gamma_r}{c_r}\frac{c_r^2-2c_r-6}{(8-c_r)(5-c_r)}u^{c_r-5}\nonumber\\ &-\frac{15}{32}B\frac{1}{(4-q)(5-2q)}\eta^{-\frac{2q}{5}}u^{2q-5}\,.
 \end{align}
On the other hand, rewriting the above expressions further in terms of $\gamma_{\dot f}$ and $c_{\dot f}$ is not so trivial in general since corrections to the frequency evolution in Eq.~\eqref{eq:f} involves two independent terms instead of one.

\section{GW Frequency Evolution From Energy Balance Law in Varying-$G$ Theories} \label{appendix_2}
In this appendix, we show an alternative approach to find $\dot f$ in varying-$G$ theories in Eq.~\eqref{eq:3.7b} by correcting and applying the energy balance law used in Ref.~\cite{Yunes:2009bv}. We begin by considering the total energy of a binary given by $E=-(G_C\mu m)/2r$. In order to calculate the leading order correction to the frequency evolution due to the time-varying gravitational constants, we  use Kepler's law to rewrite the binding energy as 
 \begin{equation}\label{eq:3.7e}
 E(f,G_C,m_1,m_2)=-\frac{1}{2}\mu (G_Cm\Omega)^{2/3}\,,
 \end{equation}
 where $\Omega=\pi f$ is the orbital angular frequency. Taking a time derivative of the above expression and using Eqs.~\eqref{eq:3.7a2}--\eqref{eq:3.7a4-2} in Eq.~\eqref{eq:3.7e}, the rate of change of the binding energy becomes
 \begin{align}\label{eq:3.7j}
 \frac{d E}{d t}=\frac{\pi^{2/3}}{6f^{1/3}G_C^{1/3}m^{4/3}}\left[-3fG_Cm(\dot{m}_{1,0}m_2+m_1\dot{m}_{2,0})\right.\nonumber\\ \left.-2m^3\eta(G_C\dot{f}+f\dot{G}_C)+m^2fG_C\eta\dot{m}\right]\,.
 \end{align}
 
We can use the following energy balance argument to derive $\dot f$. 
In GR, the time variation in the binding energy is balanced with the GW luminosity $\dot{E}_{\GW}$ emitted from the system given by
 \begin{equation}\label{eq:3.7h}
 \dot{E}_{\GW}=\frac{1}{5}G_D\left \langle\dddot{Q}_{ij}\dddot{Q}_{ij}-\frac{1}{3}(\dddot{Q}_{kk})^2\right \rangle=\frac{32}{5} r^4 G_D \mu ^2 \Omega ^6\,.
 \end{equation}
In varying-$G$ theories, there is an additional contribution $\dot E_\Gdot$ due to variations in $G$, masses, and the specific angular momentum. Namely, the binding energy is not conserved even in the absence of GW emission and the energy balance law is modified as
\begin{equation}
\label{eq:3.7g}
\frac{d E}{d t}=-\dot{E}_{\GW} + \dot E_\Gdot\,.
\end{equation}
 To estimate such an additional contribution, we rewrite the binding energy in terms of the specific angular momentum as
 \begin{equation}\label{eq:3.7f}
 E(G_C,m_1,m_2,j)=-\frac{G_C^2\, \mu \, m^2}{2 j^2}\,.
 \end{equation}
Taking the time variation of this leads to
\begin{equation}
\label{eq:EGdot}
\dot E_\Gdot=\frac{\partial E}{\partial j}\dot{j}_0+\frac{\partial E}{\partial m_1}\dot{m}_{1,0}+\frac{\partial E}{\partial m_2}\dot{m}_{2,0}+\frac{\partial E}{\partial G_C}\dot{G}_{C,0}\,,
 \end{equation}
where $\dot{j}_0$ is given by Eq.~\eqref{eq:3.7a4-3} and originates purely from the variation of $G_C$ (i.e.~no GW emission).

We are now in a position to derive the frequency evolution. 
Using Eqs.~\eqref{eq:3.7h},~\eqref{eq:3.7f} and~\eqref{eq:EGdot} in Eq.~\eqref{eq:3.7g}, one finds
 \begin{align}\label{eq:3.7i}
\frac{dE}{dt}=&-\frac{32}{5} \pi ^{10/3} f^{10/3} \eta ^2 G_C^{4/3}G_D m^{10/3}\left[1+ \frac{5 G_C^2 \eta^{3/5} m}{64G_D} \right. \nonumber\\ & \left. \times \left(\frac{\dot{m}_0}{m}+\frac{\dot{m}_{1,0}}{m_1}+\frac{\dot{m_{2,0}}}{m_2}-2\frac{\dot{j}_0}{j}+2\frac{\dot{G}_{C,0}}{G_C}\right)u^{-8}\right]\,,
 \end{align}
where $u=(\pi G_C \mathcal{M}f)^{1/3}$. Substituting this further into Eq.~\eqref{eq:3.7j} and solving for $\dot f$, one finds the frequency evolution as
 \begin{align} 
 \dot{f}=&\frac{96}{5}\pi^{8/3}G_C^{2/3}G_D\mathcal{M}^{5/3}f^{11/3}\left\{1+\frac{5}{96}\frac{G_C}{G_D} \dot{G}_{C,0} \eta^{3/5}[2m\right.\nonumber\\ 
 & \left.-5(m_{1,0}s_{1,0}+m_{2,0}s_{2,0})] u^{-8}\right\}\,,
 \end{align} 
in agreement with Eq.~\eqref{eq:3.7b}.

Along with the constancy of masses, the second term in Eq.~\eqref{eq:3.7g} was also missing in~\cite{Yunes:2009bv}. Consequently, our PPE parameters in Eqs.~\eqref{eq:beta_gdot} and~\eqref{eq:3.7d2} do not agree with Ref.~\cite{Yunes:2009bv} even when we take the limit of no time variation in masses. Difference in $\beta_{\dot G}$ is smaller than 20\% while $\alpha_{\dot G}$ differs by a factor of 7. Despite the discrepancy, we expect the projected bounds on $\dot{G}_0/G_0$ calculated in Ref.~\cite{Yunes:2009bv} to be qualitatively correct. This is because a matched filtering analysis is more sensitive to phase corrections than to amplitude ones, where the difference between our results and~\cite{Yunes:2009bv} is small.

 \bibliography{ppE}
\end{document}